# Toward Intelligent and Efficient 6G Networks: JCSC Enabled On-Purpose Machine Communications

Ping Zhang, Heng Yang, Zhiyong Feng, Yanpeng Cui, Jincheng Dai, Xiaoqi Qin, Jinglin Li, and Qixun Zhang

## Abstract

Driven by the vision of "intelligent connection of everything" toward 6G, the collective intelligence of networked machines can be fully exploited to improve system efficiency by shifting the paradigm of wireless communication design from naive maximalist approaches to intelligent value-based approaches. In this article, we propose an on-purpose machine communication framework enabled by joint communication, sensing and computation (JCSC) technology, which employs machine semantics as the interactive information flow. Naturally, there are potential technical barriers to be solved before the wide-spread adoption of on-purpose communications, including the conception of machine purpose, fast and concise networking strategy, and semantics-aware information exchange mechanism during the process of task-oriented cooperation. Hence, we discuss enabling technologies complemented by a range of open challenges. Simulation result shows that the proposed framework can significantly reduce networking overhead and improve communication efficiency.

## Introduction

As the deep integration of wireless communications and artificial intelligence, we are at the dawn of the era of ubiquitous intelligence, in which machines with onboard sensors and computing units are connected seamlessly to enable profound progress in vertical industries, including intelligent vehicular network (IVN), telemedicine, and smart manufacturing [1]. It is expected that there will be over 125 billion intelligent machines (IMs) connected to the Internet by 2030 [2], pointing at a future where IMs are the major parts of the communication grid. This trend poses tremendous challenges for the current communication system, and calls for revolutionary innovations in network design philosophy to support intelligent machine-type communications (IMTC).

In particular, the network key performance indicator (KPI) for IMTC system has contrasts and synergies with current communication system. The current communication systems have been mainly designed for content delivery, whereas the focus is to optimize the one-hop data transmission capabilities. In sharp contrast to that, sophisticated machine-based applications usually involve cooperation among groups of IMs to complete diversified tasks [3], which requires frequent and reliable sensing and control data exchange among individuals. The closed-loop communication capabilities, consisting of sensing data collection, data processing and fusion, and control data dissemination, are the focus and characteristic of IMTC optimization [4].

Under such emerging scenarios, the naive maximalist approach of blindly improving data transmission capabilities may scale up the networking complexity in terms of signaling cost and protocol overhead. We argue that the collective intelligence of networked IMs can be fully exploited to obtain situational awareness and identify the purpose of communication. In this sense, the paradigm of network design for IMTC may shift toward generating the most valuable information that can be efficiently transmitted and reconstructed at the right time instantly, to complete a specific task. Note that a certain task is the ultimate goal of a single IM or multiple IMs, and this goal is composed of different machine purposes. Therefore, IMTC would have very stringent performance requirements in *both data transmission and data processing* to support collaborative thinking and decision-making.

In this article, we employ intelligent vehicular network (IVN) [5], which is identified as one of the most complicated scenarios of IMTC, as an illustrative application that would benefit from our proposed joint communication, sensing and computation (JCSC) enabled on-purpose machine communication paradigm. Here, we provide a general view of key challenges in the future IVN.

**Whom to Communicate With:** To fully exploit the collective intelligence of IMs for task-oriented cooperation, a stable and concise topology has to be constructed to realize efficient information exchange. For example, frequent and timely information exchange is required among certain groups of vehicles for cooperative vehicle control in autonomous driving. To combat highly dynamic typology and time-varying communication environment, the ubiquitous computing and sensing capability at each vehicle should be fully exploited to enable fast networking based on the specific purpose of communication extracted from massive amounts of sensing data.

*Ping Zhang (corresponding author), Heng Yang, Zhiyong Feng (corresponding author), Yanpeng Cui, Jincheng Dai, Xiaoqi Qin, Jinglin Li, and Qixun Zhang are with Beijing University of Posts and Telecommunications, China.*

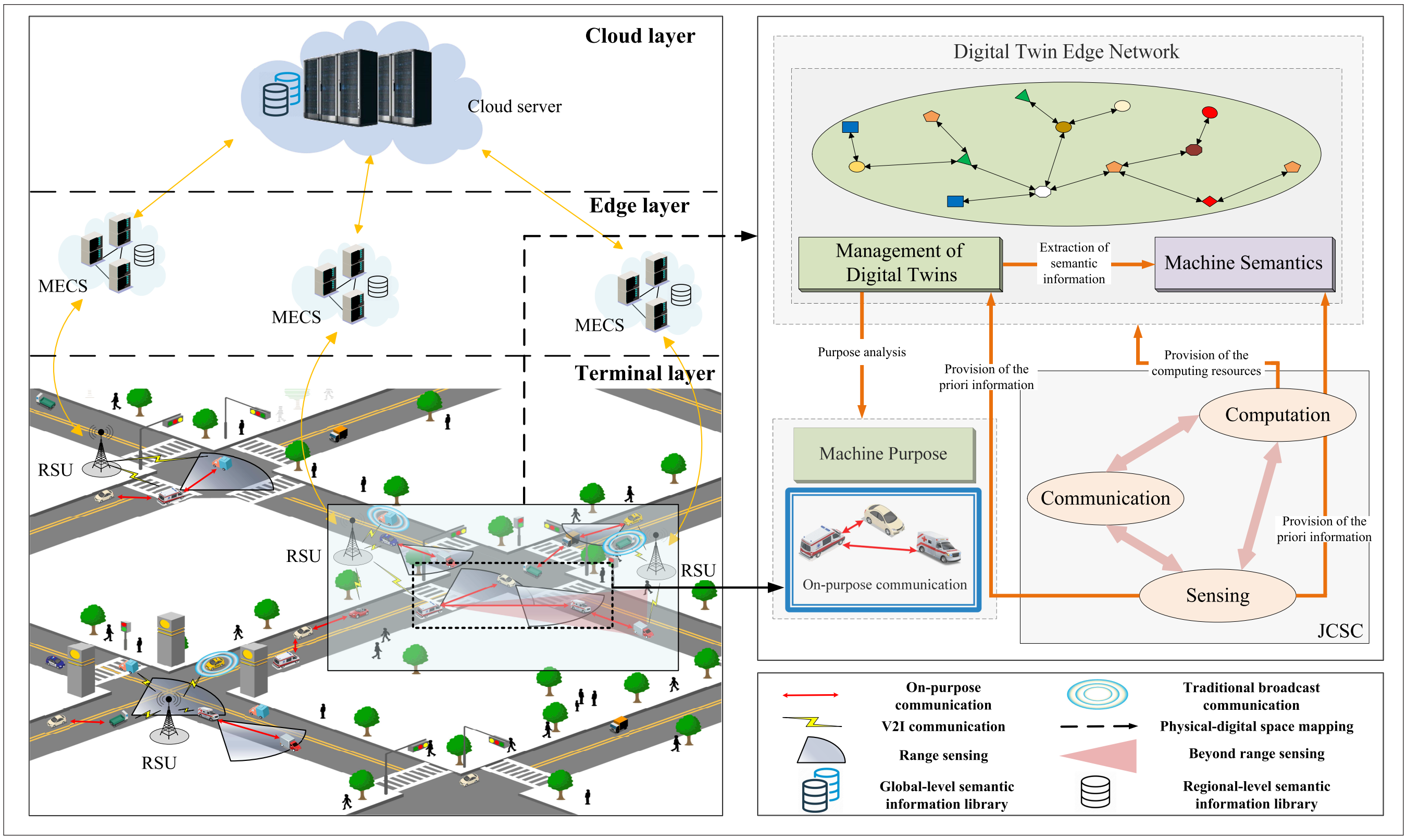

**FIGURE 1.** JCSC-based framework for on-purpose machine communications.

**How to Transmit:** Considering the emergence of collective decision making demands, blind exchange of information by broadcasting may congest the network due to limited system resources, and may affect the privacy and security of data transmission. Moreover, the highly dynamic nature of vehicular networks poses great challenges in reliable data transmission. Considering the fact that vehicles can leverage their sensing capability for target positioning, the transmitted waveform should be carefully re-designed to integrate sensing and communication functionalities to achieve secured and efficient point-to-point information exchange.

**What to Transmit:** Although the higher frequency band is proposed to be utilized in the 6G network, the spectrum resource still struggles to meet the explosive growth of data transmission demand. For example, a self-driving vehicle with ten high-resolution cameras produces two gigapixels per second [6]. The irrelevant and redundant information in sensing data leads to low system efficiency [7]. Therefore, it is of critical importance that valuable information can be extracted from raw data to realize concise and efficient information exchange.

These challenges motivate us to conceive this article on an innovative perspective for intelligent and efficient machine-type communications (MTC). Note that the technical challenges for MTC network design stem from the nature of how machines sense information and their intrinsic purpose of networking. In what follows, this article first provides a systematic introduction of IMTC systems. Then a JCSC enabled on-purpose machine communication (JCSC-OMC) framework is proposed, as shown in Fig. 1. We also give an in-depth analysis of the enabling technologies, complemented by a range of open challenges. Finally, evaluation results of our proposed framework are demonstrated.

## Intelligent Machine-Type Communications

Considering the evolution of cellular systems, the wireless communication paradigm has evolved from human-centered voice service to ubiquitous information exchange for both human-type communications (HTC) and MTC. To support MTC, 5G systems are tailored to satisfy the need for machine-type connections [8]. As the integration of artificial intelligence and wireless communication toward 6G system [1], the concept of IMTC is proposed, where the collective intelligence of multiple IMs will be leveraged to achieve precise and efficient autonomous decision making. Different from traditional HTC design, IMTC calls for ultra-reliable and low-latency closed-loop communication and high-precision positioning to enable multi-node collaboration and autonomous decision making capabilities for IMs. In this article, we propose a novel framework of on-purpose machine communications based on the following considerations.

To collaboratively accomplish specific tasks, massive sensing and control data is required to be exchanged between IMs. Assisted with JCSC technology, IMs could realize on-purpose communications by performing purpose analysis based on temporal-spatial correlations of sensing data, which can significantly reduce networking overhead and guarantee data privacy through secured directional transmission.

Equipped with sensors and computing units, IMs can obtain human-like situational awareness by utilizing machine learning methods. To efficiently realize the on-purpose communications, IMs do not need to transmit all the raw data, but only the semantic information that they have reached a consensus on, which improves the spectrum efficiency obviously.

Equipped with sensors and computing units, IMs can obtain human-like situational awareness by utilizing machine learning methods. To efficiently realize the on-purpose communications, IMs do not need to transmit all the raw data, but only the semantic information that they have reached a consensus on, which improves the spectrum efficiency obviously.

## Conception, Advantage and Framework

As discussed in our previous article [9], the new IMTC paradigm proposed in this article is based on both system theory and information theory, to optimizing the whole network. More specifically, the system entropy is introduced in this article to evaluate the orderly evolution of the JCSC-OMC network, which enables the network topology to be continuously reshaped with the time-varying wireless environment.

In view of this, to achieve multi-node collaboration and high-efficiency decision making for IMs, the JCSC-OMC network guided by system entropy is proposed in this article, aiming at realizing the on-purpose machine communication employing machine semantics as the interactive information flow, which can satisfy the ultra-reliable and low-latency closed-loop communication requirements for massive amounts of sensing, control and traffic data, and effectively improve resource utilization and communication efficiency.

### What is System Entropy?

**Definition (System Entropy):** The system entropy synthetically portrays the disorder degree of the network topology , that is, reducing the system entropy will simplify the network topology and enhance network stability. For example, the traffic system will be extremely chaotic if the behaviors of pedestrians and vehicles are not restricted, which will cause serious traffic congestion. On the contrary, the traffic efficiency will be improved once the traffic participants can be guided by the signal lights. Within this perspective, the mutual communication and sensing between vehicles and road facilities can further increase the orderly operation of traffic participants and make the IVN concise and efficient.

More specifically, the system entropy $F(S_1, S_2, ..., S_i, ...)$ is a function of the multi-level order parameters, where $S_i$ corresponds to the order parameter of the $i$-th level. It is worth noting that the order parameters of different levels are usually related to each other. From the perspective of system entropy, the key to system optimization is to improve the orderliness of the system at multiple levels. A predictable system is usually extremely well-ordered, so a direct way to improve orderliness is to make the system predict actively. By integrating machine purpose and machine semantics for intelligent computation, multiple levels of system entropy can be optimized simultaneously, resulting in the emergence of an intelligent and efficient network.

### What is Machine Purpose?

**Definition (Machine Purpose):** The machine purpose during the communication process refers to the determination and identification of the communication target. In the current communication system, such as wireless vehicular network, the omnidirectional broadcast is adopted by the transmitter to discover the communication target. The latter has to send a reply signal to the former once it receives the broadcast message, thereby establishing a communication link. In addition, it is inevitable to search forwarding nodes and transmit data through multi-hop relay when the communication target is not within the one-hop communication range. In this way, the low-latency and high-reliability demands of closed-loop communication in the IMTC scenario cannot be met by utilizing this traditional networking mode, due to the traffic congestion problem when massive amounts of information exchange between IMs, which will affect the network stability seriously. Thus, we propose an innovative paradigm, namely on-purpose machine communication. In this concept, IMs are able to transmit directional beams to whomever they want to communicate with, according to their purposes, that is, communication links can be established instantly and flexibly based on machine purpose to adapt to the dynamic changes of network topology.

For instance, the task-oriented cooperation of a certain convoy is to drive from A to B, and when this convoy is in motion, each vehicle will generate a specific purpose at each time slot. Considering a scenario that this convoy is to turn right at a certain time slot, in this case, the machine purpose of the convoy's lead vehicle is about to interact with the nodes on the right side to obtain the road status of the blind area. Specifically, the location of the optimal communication target can be determined immediately by utilizing supervised learning, unsupervised learning, and reinforcement learning methods, based on the real-time sensing data obtained by onboard sensors, and the predicted data of traffic status issued by the roadside unit (RSU). Notably, this vehicle may have multiple potential communication targets so it can calculate the system entropy when communicating with each potential communication target, and select the node with the smallest system entropy as the actual communication target. In this way, the network is the most stable and orderly.

Further, by transmitting the directional beam, this vehicle delivers data (machine semantics) directly to the communication target on the right, enabling the rapid establishment of wireless network topology on purpose to realize timely communication with low overhead, and improve the privacy and security of data transmission. Simultaneously, the vehicles on the right can repeat this process with their neighboring vehicles to obtain the traffic information at a longer distance and achieve collective intelligence of multiple IMs for over-the-horizon awareness.

### What is Machine Semantics?

**Definition (Machine Semantics):** The machine semantics refers to the information which the IMs have reached a consensus on after feature extraction and transformation from the raw data [10]. As IMs are equipped with sensors and computing units, IMs can have human-like senses, such as cognition and decision making capabilities. In the current wireless network, all the raw data (sound, video, text, etc.) needs to be transmitted between terminals. Humans should utilize the experience in their brain to extract the intrinsic semantic information behind the extrinsic observation to help understand the transmitted contents and to make correct decisions.

In contrast, since IMs can achieve human-like intelligence using artificial intelligence methods, the on-purpose machine communication paradigm enables IMs to identify the purpose-related semantic information and judiciously exchange the vital and valuable information only. Moreover, IMs and servers in different layers can continuously maintain a shared semantic information library as the grounded knowledge accumulates incrementally. In this way, compared with the current communication paradigm transmitting all raw data, the same function can be realized by primitive-concise information between machines, thus facilitating the autonomous collaboration of IMs to accomplish tasks in the complex environment.

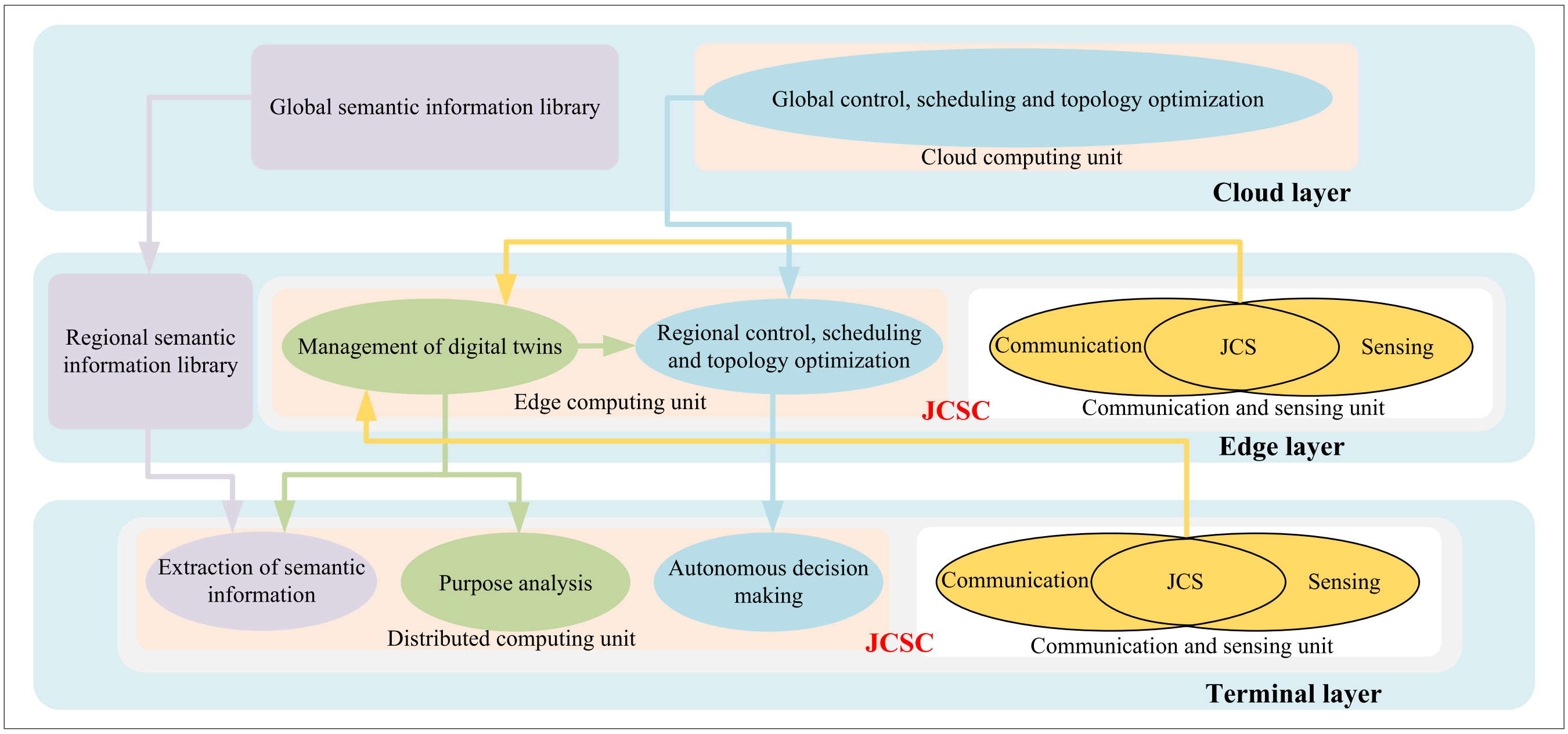

**FIGURE 2.** The function description of JCSC-OMC framework.

## JCSC-Based On-Purpose Communication Framework

To support the *on-purpose machine communication* employing *machine semantics* as the interactive information flow, we propose the JCSC-OMC framework as shown in Fig. 1. Moreover, the functions of each layer and the logical relationships between different layers are illustrated in Fig. 2.

**Terminal Layer:** The terminal layer consists of various IMs equipped with different sensing units (camera, radar, LiDAR, etc.). The raw data can be further processed based on the global (regional) semantic information library issued by the cloud (edge) sever, that is, customized features can be extracted from the raw data according to different demands. The raw data is converted into customized semantic information suitable for the specific IMs, which could realize the same functions by sending fewer bits. The location of the potential communication targets can be discovered instantly by IMs based on the sensing data. Then the directional beams are transmitted to achieve on-purpose communications once IMs generate communication purposes. In addition, the distributed intelligent computation, such as federated learning, is considered for purpose analysis, data fusion, and feature extraction to realize autonomous sensing and cognition, fast networking, and decision making for IMs by instantly encoding (decoding) semantic information at the transmitter (receiver) and performing on-purpose communications.

**Edge Layer:** The edge layer consists of base stations (BSs)[1] and mobile edge computing servers (MECSs). The BS can utilize JCSC technology to accelerate the synchronization process with the IMs within its coverage range, reducing the communication overhead and realizing the on-purpose communications. In general, there are two types of synchronization approaches utilized by BS: satellite synchronization and IEEE 1588-v2 protocol synchronization. The BS can achieve bandwidth of more than 100M and capacity of Gb/s level. Since MECSs have powerful computing ability, MECSs can receive all kinds of sensing data from IMs and BSs for regional-level data fusion to achieve regional autonomy, enhancing the robustness of the edge network. Further, MECSs are installed with the regional-level semantic information library, which is customized for specific IMs in different regions. MECS can also update the regional-level semantic information library in real-time based on the state changes and functional evolution of IMs. Also, the digital twin technology can be used to construct virtual twins of various IMs in a certain region based on the sensing data from IMs and BSs, thus completing the mapping of IMs from physical space to digital space, so as to realize the life cycle management of IMs and provide logical guidance for IMs to perform on-purpose communications.

**Cloud Layer:** The cloud layer consists of clustered servers with powerful computing ability and storage capacity. The cloud servers can aggregate and process various types of data from MECSs in different regions, hence the cloud layer can perform global autonomous scheduling and decision making based on massive data, which realizes the self-optimization, self-configuration, and self-healing of the network. Different from the regional-level semantic information library that is private for a certain region, the cloud servers are installed with the global-level semantic information library, which is used as a common library for all IMs in different regions.

## Enabling Technologies

In this section, we present a brief overview of the enabling technologies along with a range of open challenges.

[1] The RSU can be considered as a kind of BS.

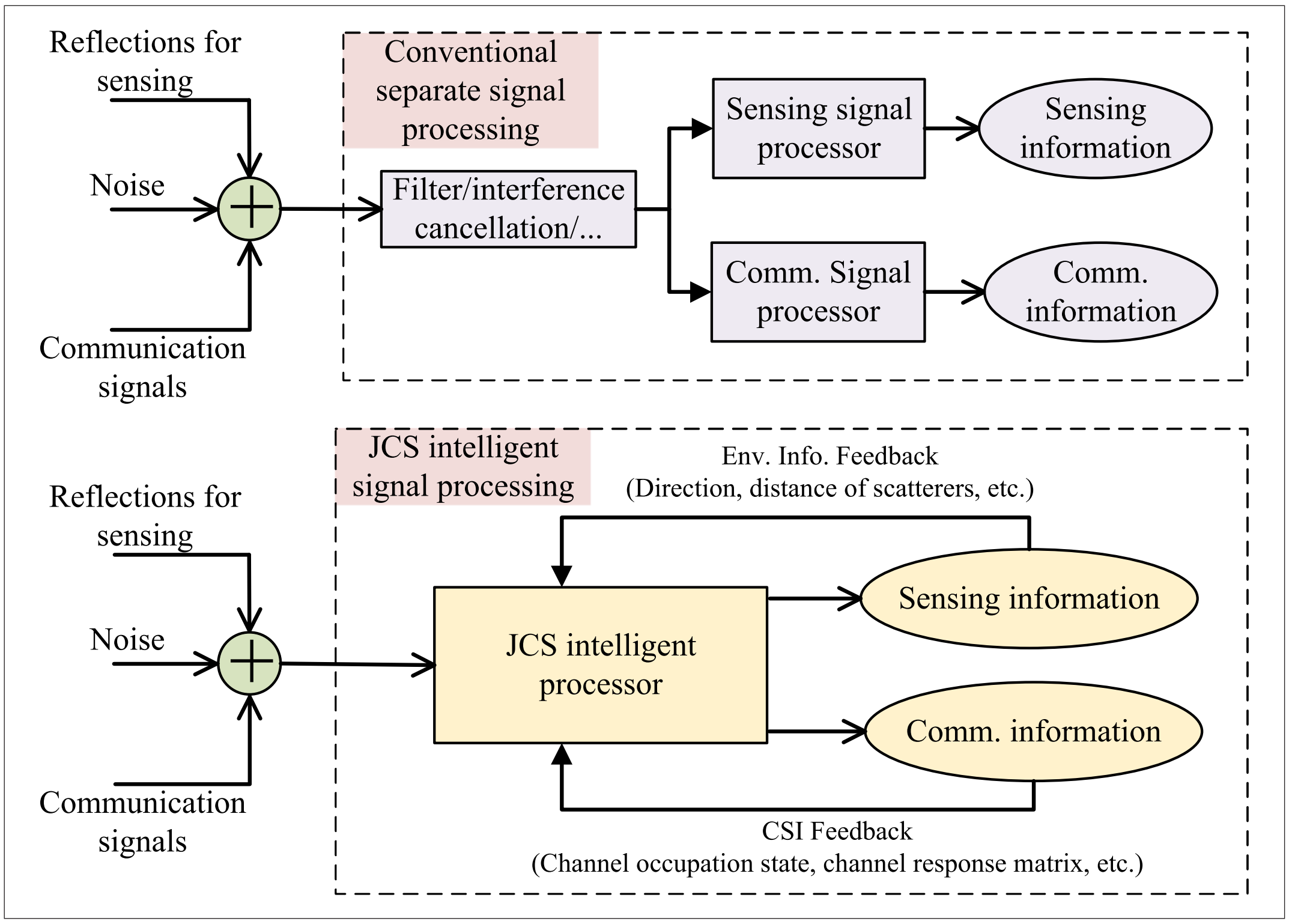

**FIGURE 3**. Intelligent signal processing enabled by JCS.

## Joint Communication, Sensing and Computation

The JCSC technology includes both the joint communication and sensing (JCS) technique and the intelligent computational technique. The state-of-the-art JCS technique utilizes the unified radio frequency (RF) transceivers and frequency band resources to achieve both wireless communication and sensing functions [11, 12]. The JCS transceiver can receive the reflection of transmitted communication beams and conduct correlation detection between the received reflection signals and the known transmit signals to obtain the motion state data that is contained in the reflection channel state information, such as the range and radial velocity of objects in the beam direction.

The comparison between conventional signal processing and intelligent JCS signal processing is shown in Fig. 3. It is highly possible that the correlation between communication channel state information and sensing data, such as the range and Doppler, can be revealed. The realization of this technique will greatly reduce the computation load and processing delay of gaining communication and sensing data. Within this perspective, intelligent JCS signal processing is the focus of future research on JCS systems and is also the focus of our future work.

The sensing data obtained additionally with the JCS technique can be taken as priori information that contains semantics for intelligent computation. The use of semantic information can significantly reduce the amount of communication data transmission, which saves more wireless resources for sensing. Artificial intelligence methods, such as supervised learning and unsupervised learning, can be used to enhance the channel estimation and prediction, which improves beamforming performance. Moreover, the reinforcement learning methods can be used to assist the beam alignment, which will greatly enhance communication reliability and power efficiency. With more reliable communication ability, the performance of collaborative computing and sensing can be further guaranteed. Besides, the intelligent computation ability not only boosts the acquisition of semantic information, but more importantly, it provides enough computing resources for the purpose analysis and generation of IMs at each time slot. In a nutshell, the JCSC technology is the key driving force of the proposed on-purpose machine communication paradigm.

**Challenge:** The surge of massive wireless terminals makes the interference problem prominent in the JCSC-OMC network. It is a key challenge to design the JCS antenna arrays and transceivers that can be adaptively used for directional communication in various spectrum bands. The potential solution is to utilize the design of hybrid analog-digital transceiver to adjust the spatial interval of virtual antenna elements of massive antenna arrays for adapting to different bands.

## Ubiquitous Computing

Ubiquitous computing intends to establish an environment full of computation and communication capabilities, emphasizing the concept of computation integrated with the environment, which provides computing resources for on-purpose communications. The key idea of ubiquitous computing is that IMs can utilize the multi-level computing resources of the cloud-edge-terminal framework optimally at anytime by intelligent and automated scheduling of these resources, which forms a computing resource sharing model. In this way, multiple IMs could utilize the computing resources from the MECSs and cloud servers on-demand to perform purpose analysis and feature extraction for semantic information in a distributed and collaborative way. Furthermore, IMs can rapidly fuse massive amounts of sensing data, and predict the trajectory of communication targets in real-time, through online training to provide priori information for on-purpose communications, enabling fast communication and autonomous decision making.

**Challenge:** The key idea of ubiquitous computing is to enable IMs to utilize the multi-level computing resources optimally at any time. Thus, how to pool the computing resources of various heterogeneous infrastructures and optimize the scheduling of computing resources is the focus of further research. The potential solution is to utilize supervised learning and unsupervised learning methods to make reasonable predictions for the usage of multi-level computing resources. Then reinforcement learning methods can be employed for efficient and rational scheduling of these computing resources.

### Digital Twin Edge Network

The digital twin is a real-time mirror of the physical entity in the digital world, which can truly reflect the operational state of the physical entity in digital space [13]. For this, with the powerful computation capacity of MECS, the digital twin edge network can be constructed in the edge layer, that is, the corresponding virtual twins are established in each region for the IMs within the coverage of the BS in this region. A mirror of the region covered by this BS is built from physical space to digital space, to realize the situation analysis and real-time decision making based on the digital space. Specifically, the regional control center in the edge layer can use the virtual twins to efficiently analyze, diagnose, simulate, and control the physical edge network based on massive data provided by the virtual twins, thus realizing the lifecycle management of the IMs and the edge network. In addition, MECS can maintain and continuously update the customized semantic information library required by an IM based on the digital data provided by the virtual twin corresponding to this type of IM. In this concept, a tightly coupled topology between digital space and physical space based on various sensing data can be constructed. Then based on this topological relationship, the IM could conduct purpose analysis rapidly and transmit data via directional beam to the targets, thus communication links can be quickly established to achieve on-purpose communications.

**Challenge:** In the JCSC-OMC network, the data is analyzed in a wide range of dimensions, including performance, capacity, energy consumption, quality, cost, efficiency, and so on. Each dimension involves a variety of different situations, making the data collection and modeling more difficult. Accordingly, how to efficiently collect such large-scale, multi-dimensional data sets, then analyze and model the collected data sets is the key issue to be addressed in the process of constructing the digital twin edge network. The potential solution is to build a unified shared data warehouse, which could simplify the data collection and modeling process.

### Sensing Assisted On-Purpose Networking

The ability of on-purpose communications will be inspired with the assistance of sensing data. The introduction of JCSC technology during the networking process can discover potential communication targets instantly, which will present the accurate network topology information for the rapid and optimal decision making based on machine purpose [14]. Thus, the transmission of beacon data and the sensing of time-varying network topology can be realized simultaneously, which significantly reduces the communication overhead and suppresses the effect of broadcast storms. Besides, the problem of hidden terminals and temporary blindness of communication can be effectively addressed. Further, wireless resources can be allocated in advance according to network status and machine purpose, accelerating the networking process of IMs. More importantly, the on-purpose selection of relay nodes and adjustment of routing schemes can be realized based on the accurate sensing of network topology, thus the JCSC-OMC network will be endowed with greater flexibility and intelligence.

**Challenge:** The priori information endures the network with a more powerful sensing ability. However, a critical issue raised in the on-purpose networking is to differentiate the identity of multiple IMs. The potential solution is to design a mapping entity between sensing and communication domains, which can provide the Internet Protocol (IP) address of the matching neighboring node with the given physical characteristics, and vice versa.

## Performance Evaluation

Based on our existing works [12, 14, 15], the orthogonal frequency division multiplexing (OFMD) waveform and the orthogonal frequency division multiple access (OFDMA) technique are utilized in the JCSC system, and we present three case studies to demonstrate the benefits of our proposed JCSC-OMC paradigm.

### Sensing-Assisted Topology Construction

IMs can obtain rich prior information about their surroundings from onboard sensors and MECS. With the aid of ubiquitous computation capabilities, IMs can perform purpose analysis and quickly discover potential communication targets. Then IMs can directly exchange data by transmitting the directional beams using JCS technique, which significantly accelerates the process of topology construction. Fig. 4 shows the required time duration for topology construction as the number of IMs increases. The sensing beam width is set as p /6. The communication radius is set as 200 m, and the network coverage area is 2000 m ¥ 80 m. As shown in Fig. 4, with the aid of prior information from sensing data, the efficiency of typology construction is greatly improved, resulting in a reduction of 49.2 percent in the number of required time slots when there are 100 nodes. Note that the omnidirectional antennas can be regarded as the directional antennas with only one beam. Thus, the discussion and analysis based on directional antennas in this article are still applicable to the omnidirectional antennas.

### Semantics-Aware Information Exchange

To support vehicle control in autonomous driving, cooperative sensing through information exchange among vehicles and RSU is leveraged to improve the precision of vehicle's situational awareness.

Considering limited system resources, the efficiency of information exchange can be improved by extracting semantic information from raw sensing data based on identified purposes. For example, images or videos captured by embedded cameras can be further processed by onboard computing units to extract features of targets (e.g., other vehicles, pavement marks). Then structural metadata or text descriptions are transmitted as auxiliary information to deliver road information. In this article, the semantic extraction network is based on Densenet backbone, which is pre-trained with ImageNet, Cityscapes, and CamVid datasets. Cityscapes and CamVid are taken from a vehicle perspective and, thus, suitable for IVN scenarios. The cropped resolution is 1024 × 512 for Cityscapes and 960 × 720 for CamVid. We use mini-batch stochastic gradient descent with a momentum of 0.95. The warmup strategy is adopted at the first 1000 iterations. As shown in Fig. 5, compared with transmitting a raw image, the spectrum efficiency is improved by ten times through semantics-aware information exchange. Note that by attending to various regions and their relations, a single inference parallelly generates a few text segments from one image that requires no more than one second.

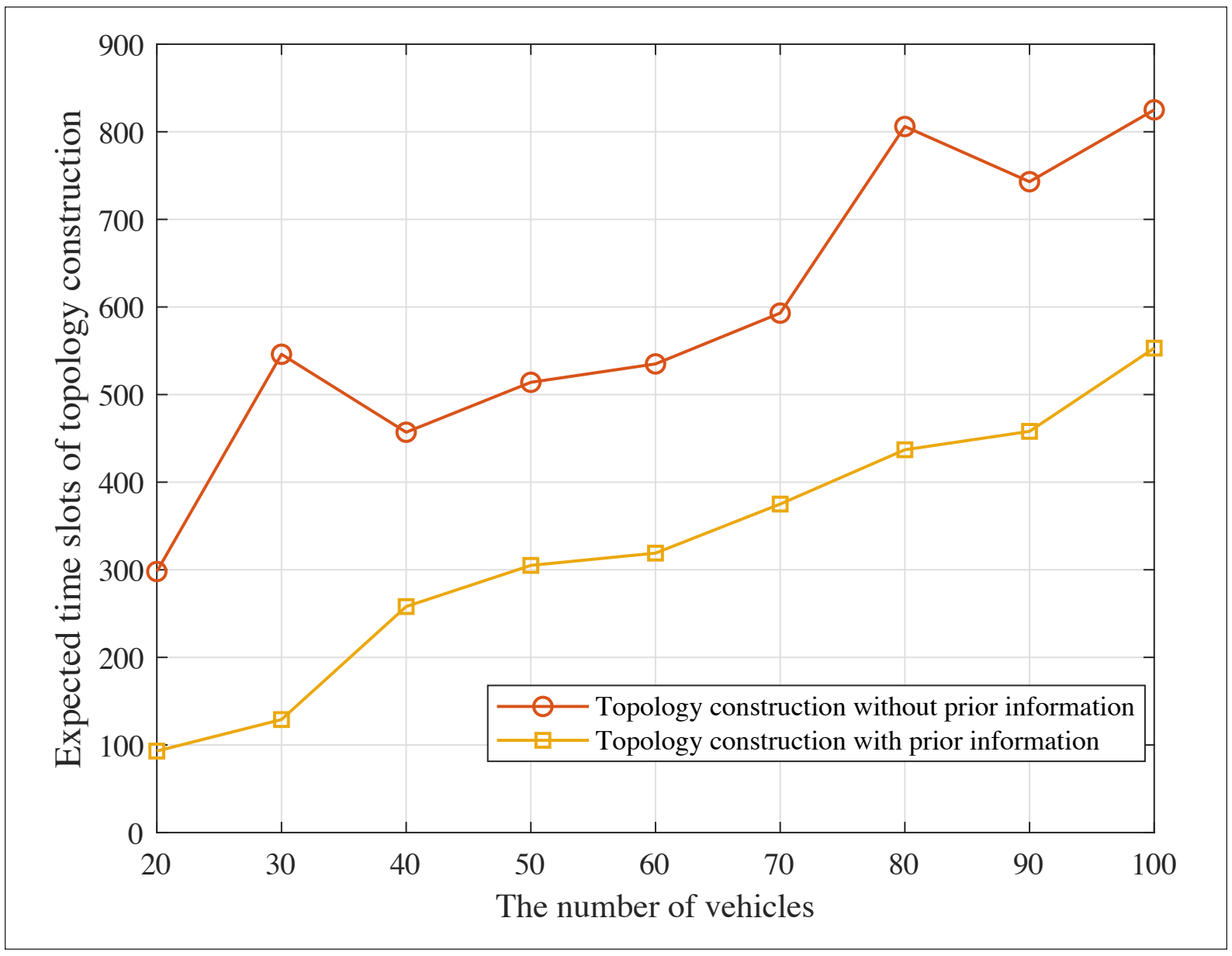

**FIGURE 4**. Time duration for topology construction.

## Semantics Enabled Sensing Capability Enhancement

The adoption of machine semantics can significantly reduce the amount of communication data transmission. Thus, in the JCSC-OMC network, more wireless resources can be used for sensing. Fig. 6 illustrates the sensing performance when transmitting different communication data volumes (corresponding to three different communication data volumes as shown in Fig. 5). Note that for each vehicle, the time-division JCS signal is adopted in the onboard JCSC system, and the radar mutual information is utilized to measure the sensing performance of this system [15]. The transmit power of onboard JCSC system is 10 W and the carrier frequency is 28 GHz with a bandwidth of 800 MHz. The antenna gain of transceiver is 18 dBi, and the transmission time of signal is 0.03 s. It can be revealed that when transmitting semantic information on purpose, compared with transmitting raw data, the sensing capability of JCSC system will be significantly enhanced, thus, providing higher quality sensing data for the IMs and RSU. Further, the IMs can quickly transmit their own sensing data in the form of machine semantics to the RSU for region-level data fusion at MECS, to achieve regional autonomy. Besides, RSU can also transmit its sensing data to IMs in the form of machine semantics to provide vehicle infrastructure cooperation services, such as over-the-horizon awareness. In this way, the real-time decision-making could be improved.

## Conclusion

In this article, we proposed a joint communication, sensing and computation enabled on-purpose machine communication framework, employing machine semantics as the interactive information flow. More specifically, we discussed relat-ed concepts, including system entropy, machine purpose, and machine semantic. We outlined the on-purpose communication framework and pinpointed its enabling technologies and challenges. Numerical results verified the feasibility and benefits of our proposed framework. We hope that our work will spur interests and open new directions for intelligent and efficient 6G networks.

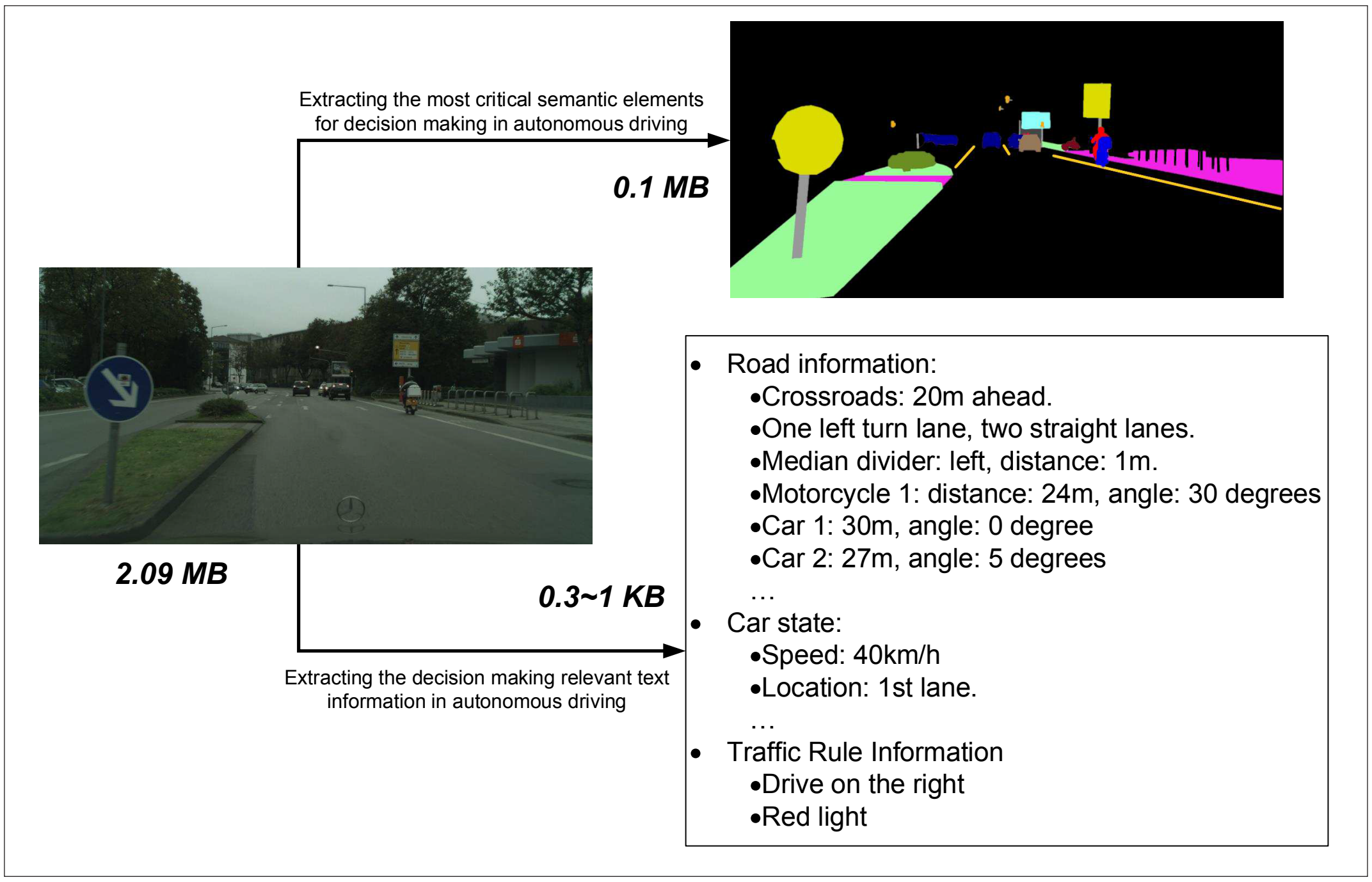

**FIGURE 5**. Raw data exchange versus semantic information exchange.

## Biographies

Ping Zhang [M'07, SM'15, F'18] is currently a professor of School of Information and Communication Engineering at Beijing University of Posts and Telecommunications, the director of State Key Laboratory of Networking and Switching Technology, a member of IMT-2020 (5G) Experts Panel, and a member of Experts Panel for China's 6G development. He served as Chief Scientist of National Basic Research Program (973 Program), an expert in Information Technology Division of National High-tech R&D program (863 Program), and a member of Consultant Committee on International Cooperation of National Natural Science Foundation of China. His research interests mainly focus on wireless communication. He is an Academician of the Chinese Academy of Engineering (CAE).

Heng Yang [S'20] received the B.Sc. degree and the M.Sc. degree from Chongqing University of Posts and Telecommunications, Chongqing, China, in 2016 and 2019. He is currently pursuing a Ph.D. degree with the School of Information and Communication Engineering, Beijing University of Posts and Telecommunications (BUPT), Beijing, China. His current research interests include radio resource management, vehicular network and joint communication and sensing network.

Zhiyong Feng [M'08, SM'15] received her B.E., M.E., and Ph.D. degrees from Beijing University of Posts and Telecommunications (BUPT), Beijing, China. She is a professor at BUPT, and the director of the Key Laboratory of the Universal Wireless Communications, Ministry of Education, P.R.China. She is a senior member of IEEE, vice chair of the Information and Communication Test Committee of the Chinese Institute of Communications (CIC). Currently, she is serving as Associate Editors-in-Chief for China Communications, and she is a technological advisor for international forum on NGMN. Her main research interests include wireless network architecture design and radio resource management in 5th generation mobile networks (5G), spectrum sensing and dynamic spectrum management in cognitive wireless networks, and joint communication and sensing system.

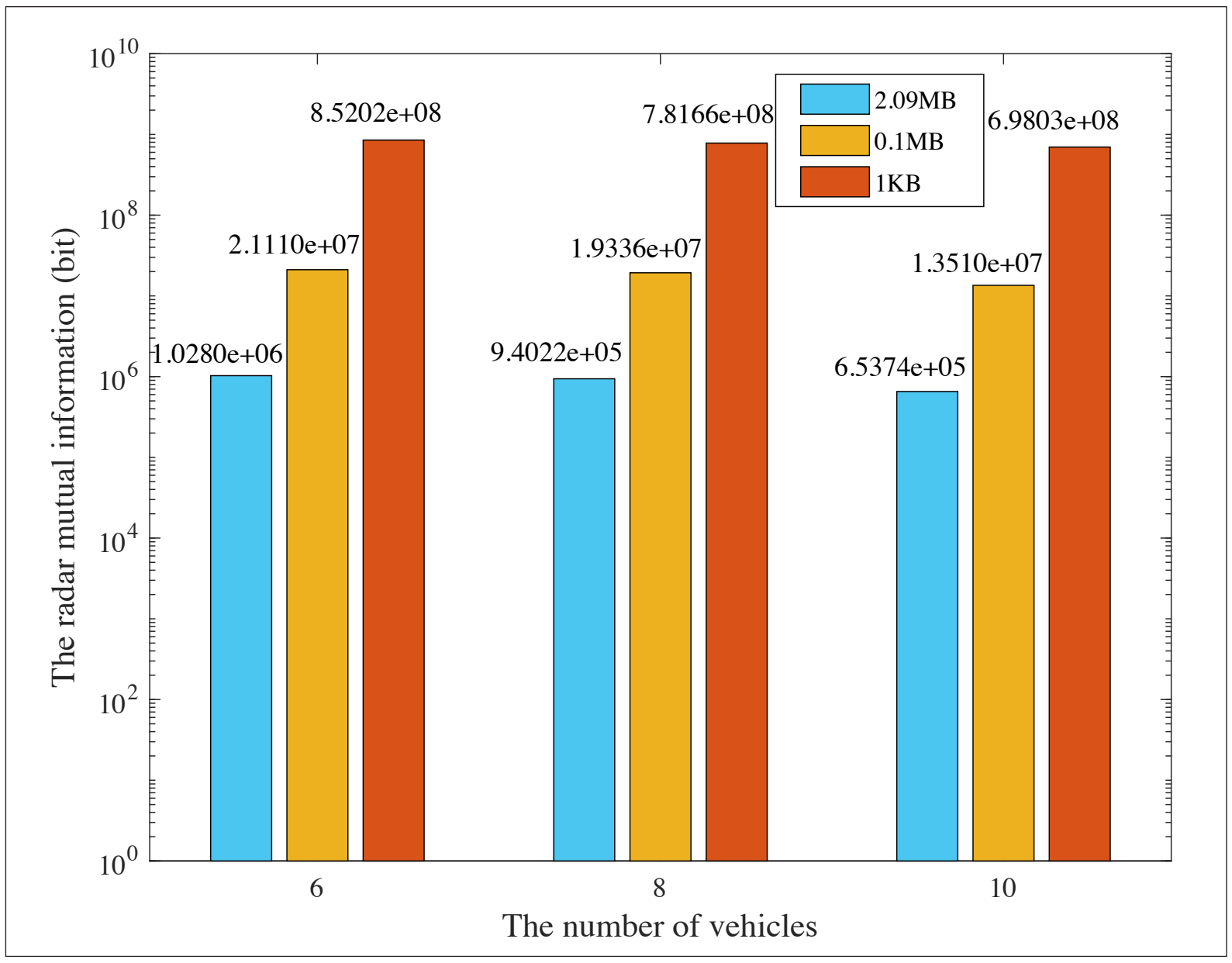

**FIGURE 6.** The sensing performance when transmitting different data volumes.

Yanpeng Cui [S'20] received the B.S. degree from the Henan University of Technology, Zhengzhou, China, in 2016, and the M.S. degree from the Xi'an University of Posts and Telecommunications, Xi'an, China, in 2020. He is currently pursuing a Ph.D. degree with the School of Information and Communication Engineering, Beijing University of Posts and Telecommunications (BUPT), Beijing, China. His current research interests include the Flying ad hoc networks, dual identity enabled networking, and integrated sensing and communication for UAV networks.

Jincheng Dai [S'17, M'21] received a B.S. and Ph.D. degree from the Beijing University of Posts and Telecommunications (BUPT), Beijing, China, in 2014 and 2019. He is currently working with the Key Laboratory of Universal Wireless Communications, Ministry of Education, Beijing University of Posts and Telecommunications. His research focuses on semantic communications, source and channel coding, machine learning for communications.

Xiaoqi Qin [S'13, M'16] received her B.S., M.S., and Ph.D. degrees from Electrical and Computer Engineering with Virginia Tech. She is currently an Associate Professor of School of Information and Communication Engineering with Beijing University of Posts and Telecommunications (BUPT). Her research focuses on exploring performance limits of next-generation wireless networks, and developing innovative solutions for intelligent and efficient machine-type communications.

Jinglin Li [M'15] received his Ph.D. degree in computer science and technology from the Beijing University of Posts and Telecommunications, in 2004. He is currently a Professor of Computer Science and Technology, and Director of the Switching and Intelligent Control Research Center (SICRC) at the State Key Laboratory of Networking and Switching Technology, China. His research interests include Internet of vehicles, intelligent machine network, and swarm intelligence.

Qixun Zhang [M'12] received a B.Eng. degree in communication engineering and a Ph.D. degree in circuit and system from Beijing University of Posts and Telecommunications (BUPT), Beijing, China, in 2006 and 2011, respectively. From March to June 2006, he was a Visiting Scholar at the University of Maryland, College Park, Maryland. From Nov. 2018 to Nov. 2019, he was a Visiting Scholar in the Electrical and Computer Engineering Department at the University of Houston, Texas. He is a Professor with the Key Laboratory of Universal Wireless Communications, Ministry of Education, the School of Information and Communication Engineering, and BUPT. His research interests include B5G/6G mobile communication system, spectrum sharing access, joint communication and sensing system for autonomous driving vehicle, mmWave communication system, cognitive radio and heterogeneous networks, game theory, and unmanned aerial vehicles (UAVs) communication. He is active in ITU-R WP5A/5C/5D standards.